\documentclass[twocolumn, tighten, numberedappendix]{aastex631}

\usepackage{color}

\usepackage{enumitem}
\usepackage{amsmath}
\usepackage{hyperref}
\usepackage{gensymb}
\usepackage{comment}
\usepackage{float}

\usepackage{tabularx}
\newcommand{\be}{\begin{equation}}
\newcommand{\ee}{\end{equation}}

\newcommand{\lcdm}{\ensuremath{\Lambda\mathrm{CDM}}}

\providecommand{\sorthelp}[1]{}

\defcitealias{nofi/etal:2025}{Paper 1}

\begin{document}
\title{Nearly Full-Sky Low-Multipole Cosmic Microwave Background Temperature Anisotropy: \\ 
II. Angular Power Spectra and Likelihood} 
\correspondingauthor{Hayley C. Nofi}
\email{hnofi1@jh.edu}

\author[0000-0001-9694-1718]{Hayley~C.~Nofi}
\affiliation{The William H. Miller III Department of Physics and Astronomy,\\ Johns Hopkins University, 3400 N. Charles Street, Baltimore, MD 21218 USA}

\author[0000-0002-2147-2248]{Graeme E. Addison}
\affiliation{The William H. Miller III Department of Physics and Astronomy,\\ Johns Hopkins University, 3400 N. Charles Street, Baltimore, MD 21218 USA}

\author[0000-0001-8839-7206]{Charles L. Bennett}
\affiliation{The William H. Miller III Department of Physics and Astronomy,\\ Johns Hopkins University, 3400 N. Charles Street, Baltimore, MD 21218 USA}

\author[0000-0001-9054-1414]{Laura Herold}
\affiliation{The William H. Miller III Department of Physics and Astronomy,\\ Johns Hopkins University, 3400 N. Charles Street, Baltimore, MD 21218 USA}

\author[0000-0003-3017-3474]{J. L. Weiland}   
\affiliation{The William H. Miller III Department of Physics and Astronomy,\\ Johns Hopkins University, 3400 N. Charles Street, Baltimore, MD 21218 USA}

\shortauthors{Nofi et al} 
\shorttitle{}

\begin{abstract}
\noindent
We present a CMB temperature power spectrum measurement at large angular scales from WMAP and Planck maps that were cleaned of foregrounds using a template-based approach described in the first paper of this series. We recover essentially the full-sky CMB information at multipoles $\ell<30$ with only 1\% of pixels near the Galactic plane masked and no inpainting. Notable features continue to appear: (a) a low quadrupole power compared to the best-fit Planck 2018 $\Lambda$CDM spectrum at $2.2\sigma$, (b) a dip in the range $20 \le \ell \le 27$, and
(c) an overall $\ell<30$ power level low of the $\Lambda$CDM prediction derived from higher multipole moments. Given the different methodology from previous analyses and the nearly full-sky solution presented here, these features do not plausibly arise from foreground contamination, systematic errors, masking, or mode-mixing. Our overall $\ell<30$ amplitude constraint is consistent with published WMAP (77\% sky fraction) and Planck (86\%) results at $1.2\sigma$ and $0.6\sigma$, respectively, accounting for the improvement in statistical precision. We present a new $\ell<30$ likelihood for use with the \texttt{Cobaya} package. Parameter constraints from joint fits with the higher-multipole Planck data are consistent with the published Planck results, for example we find $H_0=67.35\pm0.54$~km~s$^{-1}$~Mpc$^{-1}$ in a joint \lcdm\ fit.

\end{abstract}
\keywords{\href{http://astrothesaurus.org/uat/322}{Cosmic microwave background radiation (322)}}

\section{Introduction}
\label{sec:intro}

The largest angular-scale cosmic microwave background (CMB) temperature modes on the sky have been measured by the COBE
\citep{bennett/etal:1996, hinshaw/etal:1996a}, WMAP \citep{bennett/etal:2013, hinshaw/etal:2013}, and Planck \citep{planck/05:2018, npipe:2020} space missions. In all three cases, significant sky cuts were applied to avoid the adverse effects of contamination by strong foreground emission. 

The space missions have found some oddities in the CMB temperature measurements that have attracted attention over the years. For example, in the temperature (TT) spectra: (1) the quadrupole ($\ell=2$) amplitude is low compared to \lcdm\ model predictions, (2) there is a noticeable feature near multipole moment $\ell=22$, and (3) taken as a whole the $\ell<30$ portion of the power spectrum lies below the level of the $\Lambda$CDM prediction based on $\ell \geq 30$. In addition, there are other large-scale oddities commonly called CMB anomalies \citep[e.g.,][]{bennett/etal:2011, schwarz/etal:2016, planck/07:2018}.  

The significance of $\ell<30$ CMB anomalies, and other uses of this portion of the angular power spectrum, is affected by masking and Galactic emission. Foreground cuts result in the loss of CMB information, disrupting the orthogonality of the spherical harmonic basis functions and causing mode mixing.

The goal of this paper is to make a new measurement of the $\ell<30$ temperature angular power spectrum with only 1\% of pixels masked and with no inpainting. We start with the foreground cleaned maps from \cite{nofi/etal:2025}. The new CMB maps provide an opportunity to avoid any significant mode-mixing and recover essentially complete CMB information. We focus on $\ell<30$, matching the Planck Collaboration's choice of multipole split for the low-$\ell$ portion of the temperature power spectrum. Information from smaller angular scales is still present in the maps and power spectra, being ultimately limited by the $1^\circ$ map resolution.

This paper is organized as follows.
In Section~\ref{sec:data}, we provide a summary description of the CMB maps from \citealt{nofi/etal:2025} that we use in our power spectrum analysis. Section~\ref{sec:analysis} provides details of the power spectrum estimation, associated uncertainties and comparison with previous results in the literature.  We discuss our conclusions in Section~\ref{sec:conclusions}.

\section{Observations and Data}
\label{sec:data}

\cite{nofi/etal:2025} use a foreground template-cleaning method to derive nearly full-sky CMB maps at four frequencies (70, 94, 100 and 143 GHz) that were particularly well cleaned, based on map differences, power spectra differences, histogram comparisons, and other statistical comparisons. The rms of the cleaned difference maps for the 1\% mask are between 3.3-9.1\% of the 70 $\mu$K CMB rms (Planck best-fit $\Lambda$CDM model), from Section 3.7 of \cite{nofi/etal:2025}. For these frequencies, the cleaning process is aided by the diffuse Galactic foreground emission being near its minimum contribution to the total sky signal. The foreground template removal is accomplished by fitting Galactic foreground templates to individual frequency maps without regard to emission component separation, and with a mask that excludes only 1\% of map pixels from the fit.  The masking
removes bright pixels where the four cleaned frequency maps are in least agreement.  In this respect, the cleaning method
differs from full-sky CMB estimators that have utilized inpainting (e.g., {\texttt{SMICA}}, see \citealt{planck/04:2018}) to mitigate the effects of difficult-to-clean sky regions.

\begin{figure}[ht]
    \centering
    \includegraphics[width=3.3in]{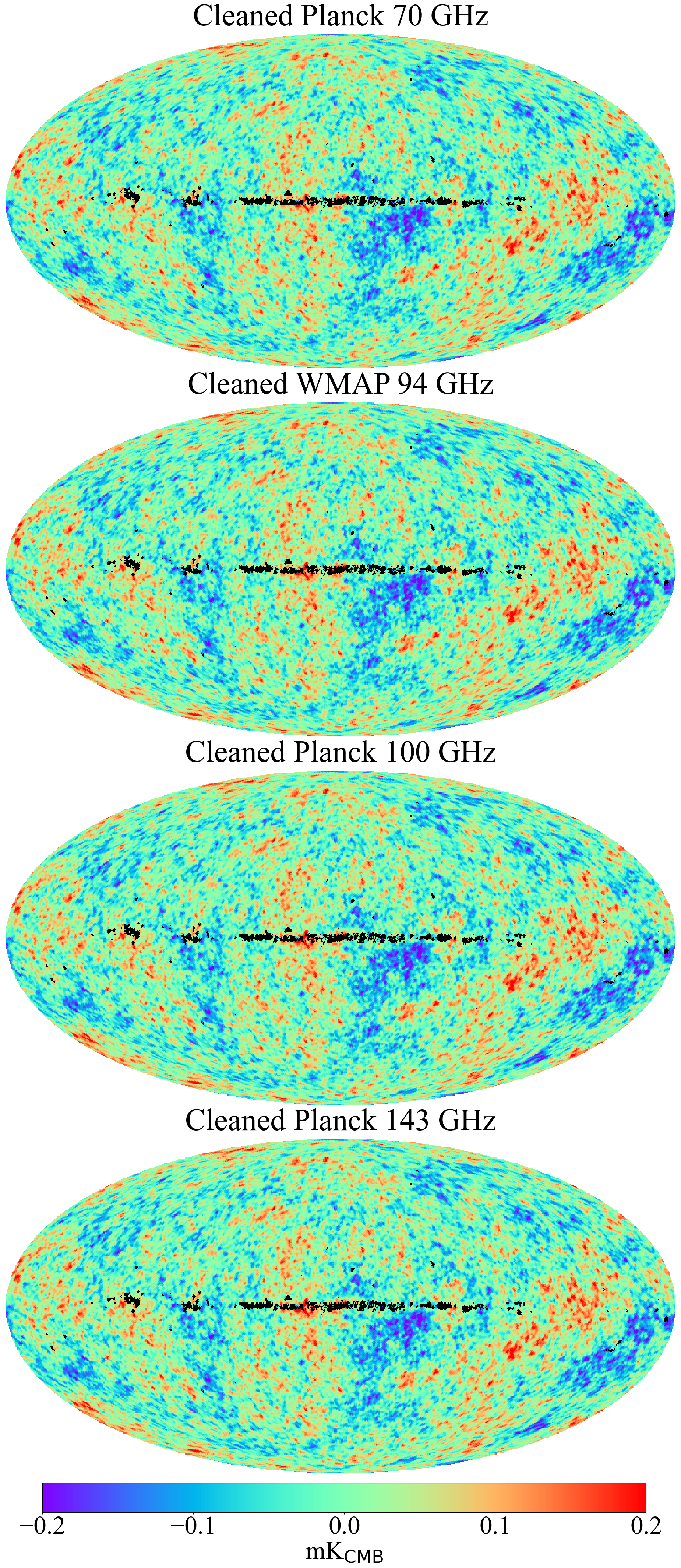}
    \caption{The CMB maps that are used to form the power spectra in this work. The 1\% of pixels that are
    masked are in black.}
    \label{fig:masked_cln_maps}
\end{figure}

The four CMB maps are shown in Figure~\ref{fig:masked_cln_maps}, with the masked pixels in black (primarily close to the Galactic plane).
The three CMB maps at 70, 100 and 143 GHz are foreground-cleaned Planck PR3 \citep{planck/01:2018} T maps, and the 94 GHz map
is a foreground-cleaned version of the WMAP 9yr, deconvolved, coadded, and smoothed W-band temperature map 
\citep{bennett/etal:2013}.  All four maps have been smoothed to a common resolution of $1\degree$ FWHM and
degraded to HEALPix\footnote{\url{http://healpix.sf.net}} $N_{\rm{side}}=128$ (pixels $\sim27'$ on a side).  We find no significant
differences in our results if we use SRoll2-processed Planck maps \citep{delouis/etal:2019} in place of PR3.  We do not use
PR4 maps due to the presence of an intentional residual zodiacal light component \citep{npipe:2020}.

\section{Data Analysis}
\label{sec:analysis}

\subsection{Power Spectrum Estimation}
\label{sec:power_spectra}

To minimize potential bias resulting from foreground residuals and instrumental noise in the individual frequency
CMB maps, 
we do not utilize auto-spectra, but rather limit our computations to 
cross spectra between
the CMB maps at the four frequencies. The cross-spectra are designated as $70\times94$, $70\times100$, $70\times143$, 
$94\times100$, $94\times143$, and $100\times143$. 
Cross-spectra are evaluated using the HEALPy \citep{healpy:2019} implementation of {\texttt{anafast}} with the 1\% mask of  \cite{nofi/etal:2025} applied to each of the two input CMB maps of differing frequencies. 

Setting 1\% of pixels to zero biases the power spectrum estimate low by around 1\%, corresponding to around a $0.2\sigma$ bias on the overall $\ell<30$ spectrum amplitude. We correct for this using the mode-coupling matrix method described by \cite{hivon/etal:2002}. This approach returns an unbiased spectrum, but generally with larger uncertainties than, for instance, quadratic maximum-likelihood estimators \citep[e.g.,][]{tegmark:1997}. For such minimal masking, however, simulations show that the uncertainties in the corrected spectra at $\ell<30$ remain effectively indistinguishable from the full-sky cosmic variance expectation. We finally correct the spectra for beam ($1^\circ$ FWHM) and $N_{\rm{side}}=128$ pixel window functions.

We also investigate the potential for bias in the derived CMB power spectra resulting from chance correlations between the
CMB and the templates used in \cite{nofi/etal:2025} to fit out the Galactic foreground emission.  We discuss this topic in greater 
detail in the next section.

\begin{figure*}[ht]
    \centering
    \includegraphics[width=6.5in]{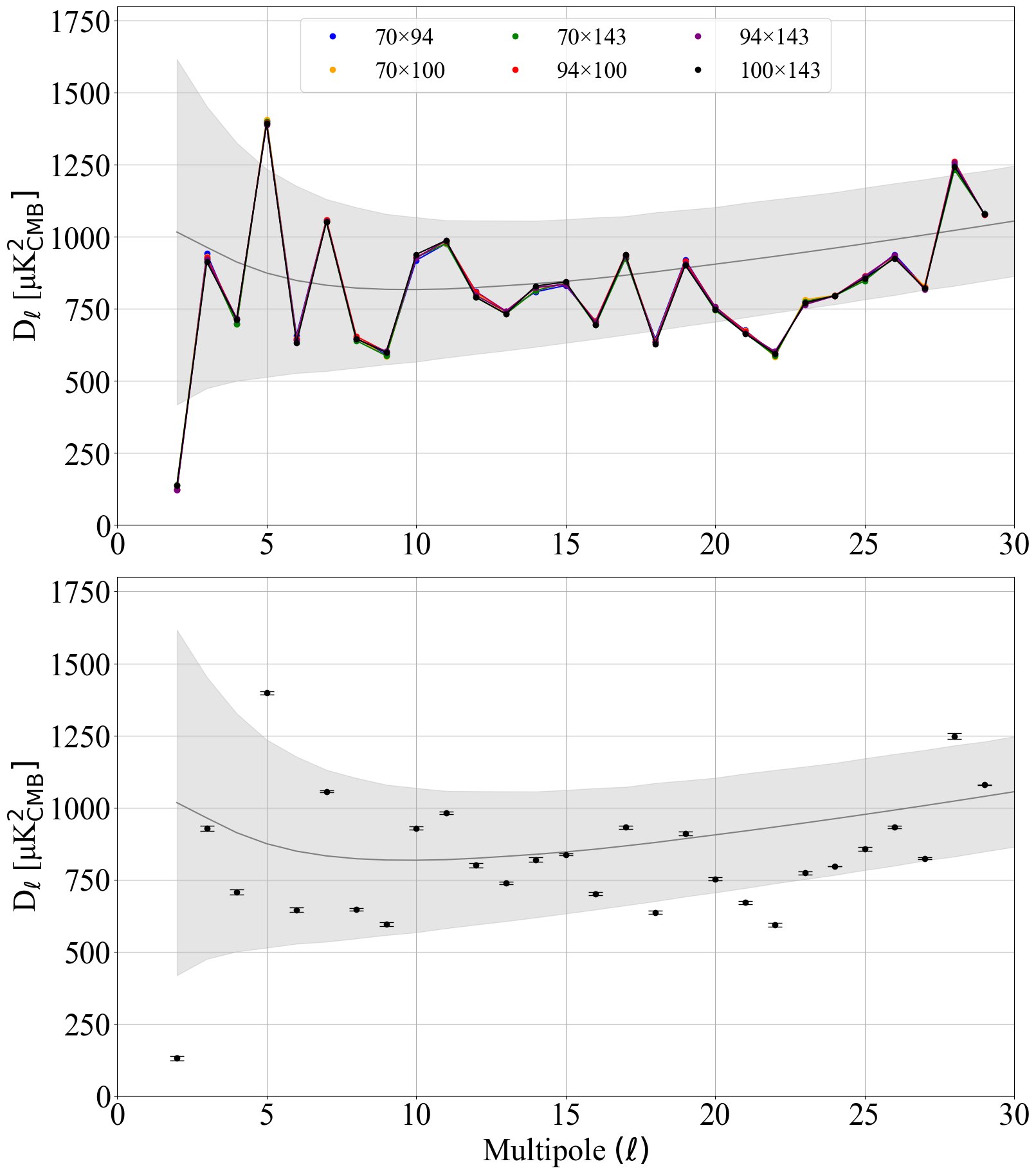}
    \caption{Cross-spectra pairs of the four best cleaned maps from \cite{nofi/etal:2025} with mask correction.
    The gray line is the 2018 Planck baseline \lcdm\ model,
    with the cosmic variance $1\sigma$ band in gray. The top plot shows the individual pairs with connecting lines added to show differences. The bottom plot shows the mean of the cross-spectra pairs with the standard deviations.}
    \label{fig:cross_spectra}
\end{figure*}

\begin{figure*}[ht]
    \centering
    \includegraphics[width=6.5in]{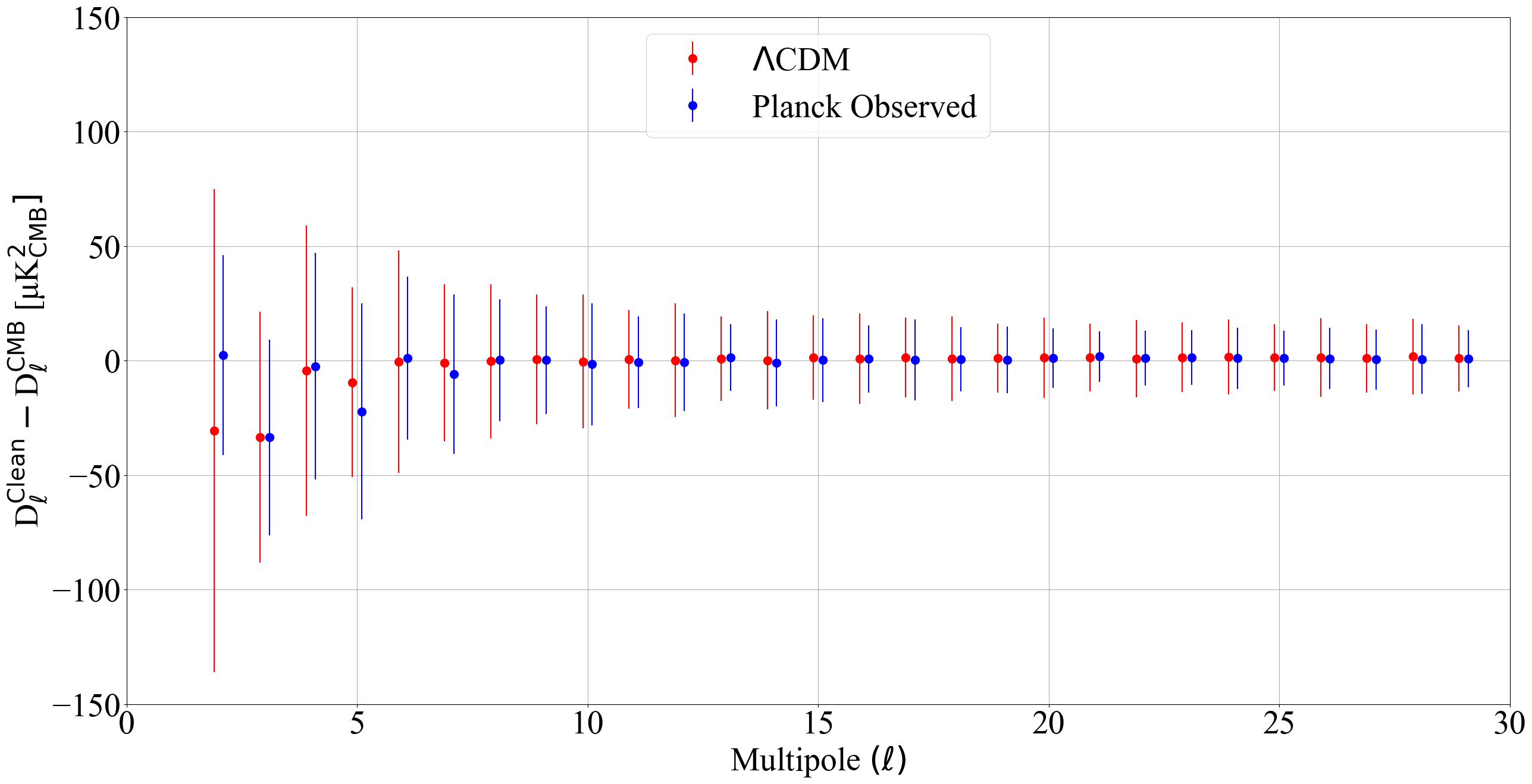}
    \caption{The mean bias estimates for the cross-spectra pairs of the four best cleaned maps. To calculate the bias, we generate 10,000 simulated CMB maps based on either the $\Lambda$CDM model or the Planck Observed sky. Foreground models for each of the four frequencies are applied to the simulated maps based on the fitting procedure described in \cite{nofi/etal:2025}. The simulated maps are cleaned using the same method as applied on the real sky. The bias is the difference between the cleaned simulated CMB maps and pure simulated CMB maps. We do not apply a bias correction to the spectra in this paper, as discussed in the text.}
    \label{fig:chance_corr_bias}
\end{figure*}

\begin{figure}[ht]
    \centering
    \includegraphics[width=3.4in]{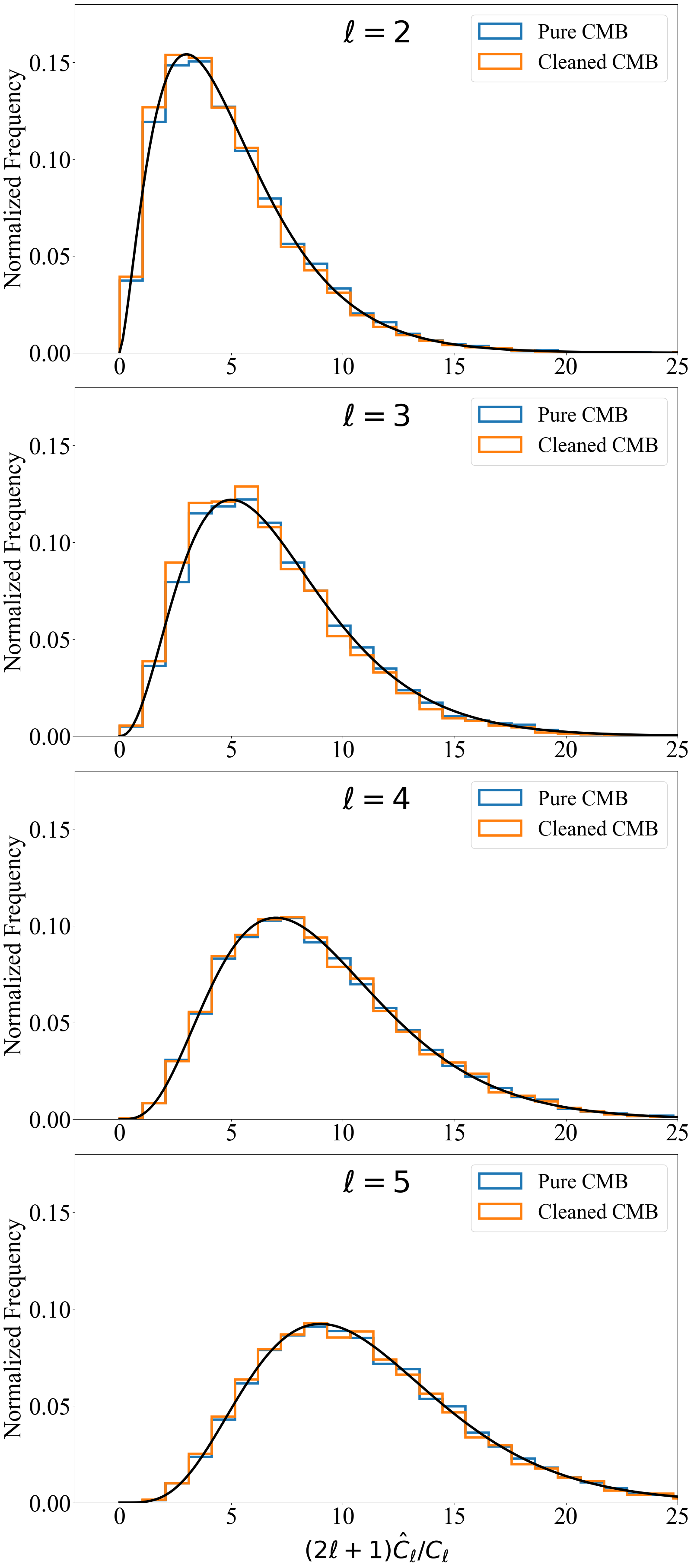}
    \caption{The histograms of the pure CMB auto-spectra and the mean of the four cleaned CMB auto-spectra for 10,000 \lcdm\ simulations. The black line is the
    $\chi^2$ distribution for the $\ell$-mode in each subplot. The pure CMB and the cleaned CMB are well fit by the 
    $\chi^2$ distribution and are very similar for each $\ell$-mode without applying a bias correction.}
    \label{fig:chisquare_hist}
\end{figure}

Resultant mask-corrected cross-spectra are shown in Figure~\ref{fig:cross_spectra}.   In the top panel,
each of the six individual cross-spectra are shown as a
different colored line.  Technically, using a line to connect the individual multipoles is misleading, since there
is no information between integer multipoles.  However, the lines help the eye discern the spread between the six cross-spectra at each multipole. We use this spread as one measure of uncertainty
in the power spectrum estimation.  In the bottom panel, the points represent the flat-weighted mean of the six cross-spectra, with uncertainties 
computed as the standard deviation of the six spectra at each monopole.
The gray band delineates the $1\sigma$ cosmic variance uncertainty for the model.
The theory spectrum, shown in gray, is the baseline \lcdm\ model from the Planck 2018 release{\footnote{COM\_PowerSpect\_CMB-base-plikHM-TTTEEE-lowl-lowE-lensing-minimum-theory\_R3.01.txt}}. In the Introduction, we mentioned a number of
features seen in the low-$\ell$ TT spectra from previous studies of WMAP and Planck data.  Our spectra exhibit these features as well, with the most visually obvious being the low quadrupole value. We discuss these features later in this section.

\begin{figure*}[ht]
    \centering
    \includegraphics[width=6.5in]{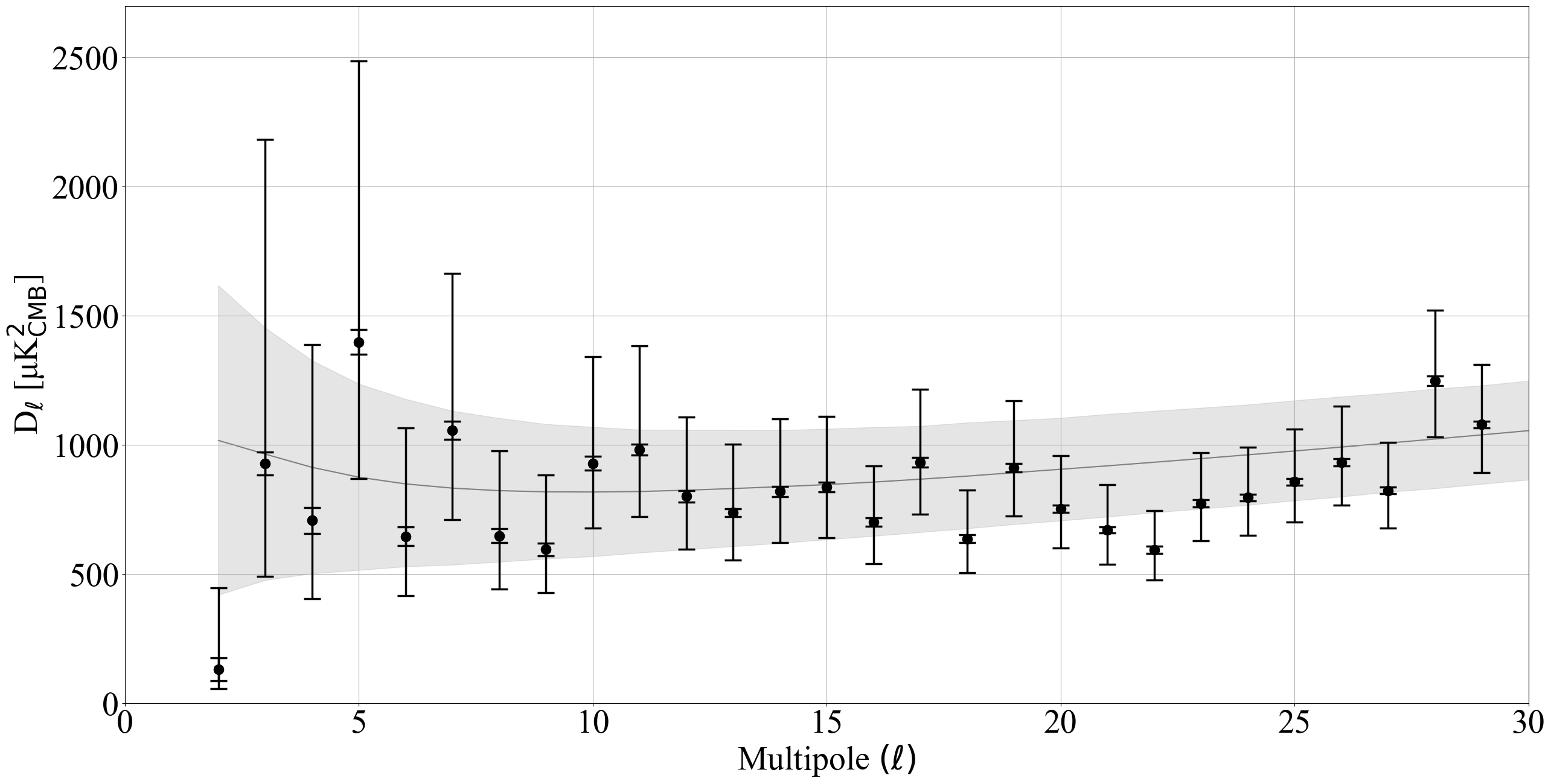}
    \caption{The mean of the cross-spectra of the four best cleaned maps over 99\% of sky is shown. The gray curve is Planck 2018 \lcdm\ spectrum and the gray band is the $68\%$ cosmic variance level. The maximum likelihood points are shown from this work with inner error bars from the quadrature sum of the foreground cleaning errors, shown in the bottom plot of Figure \ref{fig:cross_spectra}, and the bias uncertainty from the simulated CMB maps based on the Planck Observed sky shown in Figure \ref{fig:chance_corr_bias}. The outer error bars include the cosmic variance uncertainty computed from the inverse-gamma posterior distribution.}
    \label{fig:cross_spectra_mean_errors}
\end{figure*}

\begin{figure*}[ht]
    \centering
    \includegraphics[width=6.5in]{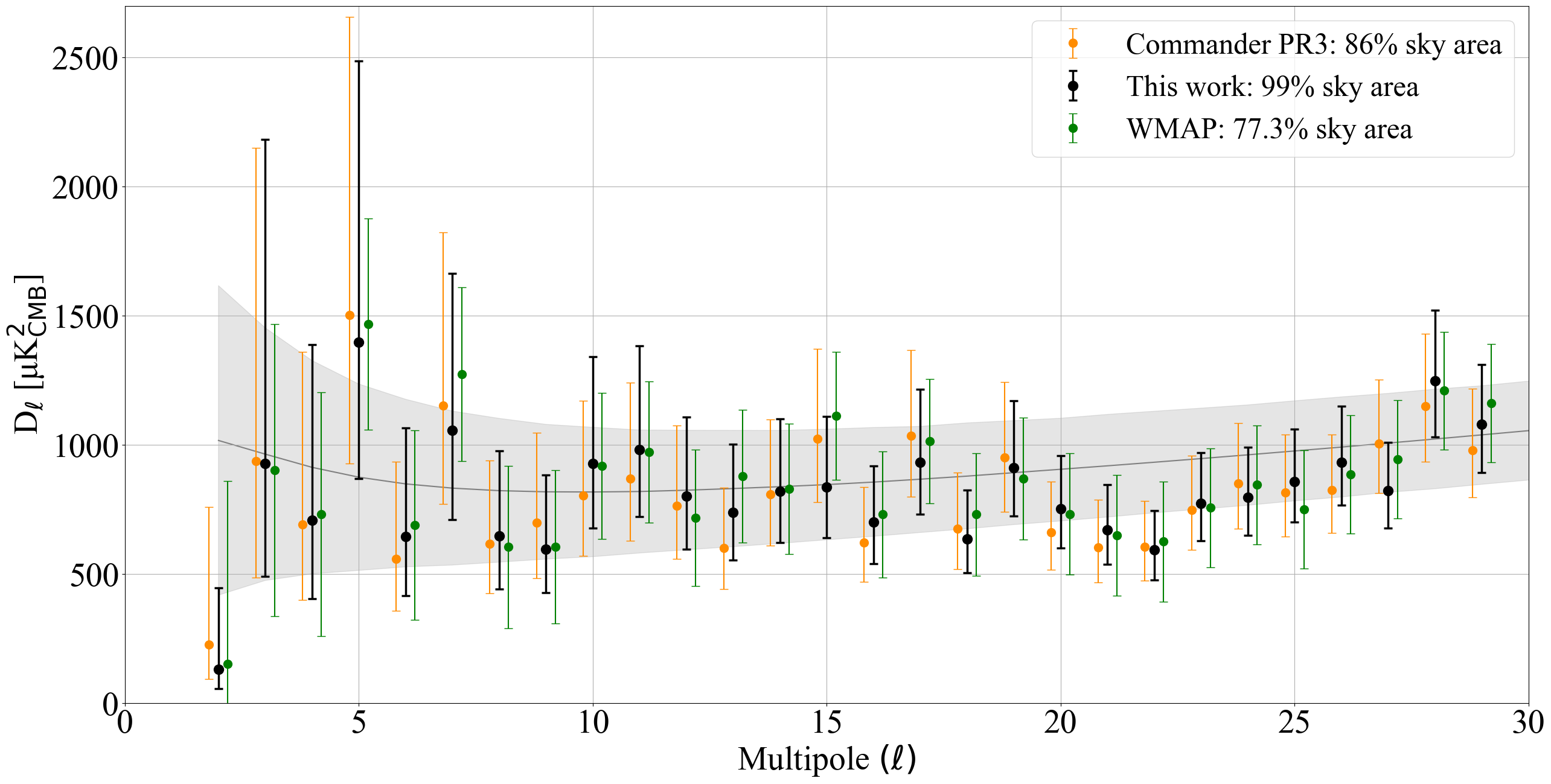}
    \caption{A comparison of the power spectra from this work (black), the Planck low-$\ell$ power spectrum (orange), and the WMAP low-$\ell$ power spectrum (green). The gray line is Planck 2018 \lcdm\ spectrum and the gray band is the $68\%$ cosmic variance level. The three spectra were derived with different analysis methods, and the sky areas are also different, meaning the CMB sample variance is not identical, so precise agreement is not expected. The Planck and WMAP spectra are plotted as provided from their public releases. The error bars for this work are  the cosmic variance uncertainties shown in Figure \ref{fig:cross_spectra_mean_errors}. Despite differences in methodology and sky coverage, there is clear similarity between the three power spectra.}
    \label{fig:cross_spectra_overplot}
\end{figure*}

\subsection{Chance Correlation with Foreground Templates}
\label{sec:uncertainties}

In this section we discuss chance correlations between the CMB fluctuations and the foreground cleaning templates. Based on simulations we have performed, these correlations have a negligible effect for all but the very lowest multipoles ($\ell\lesssim5$), and we ultimately decided not to make any adjustment to our measured spectra or $\ell<30$ likelihood to account for them. However, tests or calculations that are especially sensitive to the lowest multipoles may be more impacted, and the simulations can be used to quantify this.

At each target frequency, \cite{nofi/etal:2025} restricted the determination of the foreground template fitting coefficients to data within Galactic latitudes $\vert b \vert\leq 10^\circ$. This choice was intended, in part, to help mitigate chance correlations between the CMB signal and the foreground templates by using a large sky region where the foreground emission is at its strongest, however it does not eliminate the effect entirely.

We use two sets of 10000 CMB temperature map realizations in conjunction with a known input foreground model to examine the impact of the chance correlations. One set is generated using the same baseline Planck 2018 \lcdm\ theory spectrum that is shown in Figure~\ref{fig:cross_spectra}. For the second set, we wanted to avoid making any particular theoretical model choice, and so constructed the simulations to contain the same CMB power spectrum as that derived from Planck PR3
observations\footnote{COM\_PowerSpect\_CMB-TT-full\_R3.01.txt}, including, for example, the lower quadrupole power than \lcdm. We used the HEALPy {\texttt{synalm}} routine to generate random $a_{\ell m}$, but then rescaled the $a_{\ell m}$ for each realization such that the simulated power given by $\frac{1}{2\ell+1}\sum_m|a_{\ell m}|^2$ matched the Planck observed value at each $\ell$. For a given $\ell$, different realizations then have the same power, but the distribution of power between different $m$-modes corresponding to this $\ell$ is still randomized. This simulation set is labeled `Planck Observed'.

The input Galactic foreground model at each frequency is taken to be the same weighted linear combination of foreground templates used in \cite{nofi/etal:2025}
to clean the frequency maps. The simulated sky realization at each frequency is then the sum of the (fixed) Galactic foreground model and the CMB realization. Each map realization is then template cleaned using the same procedure as described in \cite{nofi/etal:2025}, and the six cross-power spectra combinations are evaluated with the 1$\%$ analysis mask. 

Figure~\ref{fig:chance_corr_bias} shows the mean and standard deviation of the difference between the cleaned and original pure CMB power spectra from both sets of simulations. In each case we use the mean of the six cross-spectra as our best power spectrum estimate, as for the real data. In general the spread in the cleaned-minus-original difference is smaller for the Planck Observed simulations. For fixed templates the size of the chance correlation scales with the square root of the input CMB power, and the measured power generally falls low of the \lcdm\ best-fit model, particularly at $\ell=2$ and $\ell\simeq20-25$. For $\ell=2$, in particular, the spread in power from the chance correlation is a significant fraction of the measured power. Since the error estimates scale with the power, and the Planck Observed power is lower overall than the \lcdm\ model for $\ell < 30$, the Planck Observed uncertainties are lower relative to those of the model.

If the chance correlation was a random effect that was unrelated to the CMB signal we could attempt to correct for it in a single realization by subtracting the mean values for one of the sets of points shown in Figure~\ref{fig:chance_corr_bias}. However, we find that applying a bias correction to the cleaned spectra in this way does not in fact lead to better recovery of any quantities derived from the power spectrum. 

As an example, we considered rescaling the true input theory power by an overall multiplicative amplitude parameter for the \lcdm\ simulation set, and checking the consistency of this amplitude with the expected value of 1.0 using only $\ell=2, 3, 4, 5$. More details of this amplitude fitting are provided in Section~\ref{subseclikelihood}. We obtained mean amplitudes of 0.999 for the pure CMB, 0.982 for the cleaned CMB, and 1.023 for the bias-corrected cleaned CMB spectra. The power spectrum bias correction over-corrects for the amplitude discrepancy. The standard deviation of the amplitudes (reflecting the cosmic variance spread) was around 0.25, so the effect of the chance correlation is subdominant regardless, however this illustrates that there is some non-random coupling between the input CMB power and how the power is redistributed by the chance correlation.

While Figure~\ref{fig:chance_corr_bias} shows that individual realizations can exhibit sizable shifts in power at low multipoles, we find that the overall distribution of the measured power is minimally impacted. This is illustrated in Figure~\ref{fig:chisquare_hist}. We plot the histogram of the normalized quantity $(2\ell+1)\hat{C}_{\ell}/C_{\ell}$, where $\hat{C}_{\ell}$ is the power measured in the simulations, and $C_{\ell}$ is the input theory. For full-sky, pure CMB fluctuations this quantity is distributed as a $\chi^2$ random variable with $2\ell+1$ degrees of freedom, and the corresponding probability density function is overplotted on each panel of the Figure. There is excellent agreement between the pure CMB and the cleaned CMB distributions, even for $\ell=2,3$.

Calculations using the wider $\ell<30$ multipole range, where the lowest few multipoles have limited relative weight, show negligible dependence on the chance correlations. For example, applying the overall amplitude test described above to $2\leq\ell\leq29$ yields mean amplitudes of 1.0026 for the pure CMB simulations, and 1.0030 for the cleaned simulations, to be compared with a statistical cosmic variance scatter of 0.048. Motivated by this, in the absence of a simple and effective way to implement a bias correction, we do not correct for the chance correlations in the power spectra or the $\ell<30$ likelihood described in Section~\ref{subseclikelihood}, below.

Some of the CMB anomaly tests that have been discussed in the literature (e.g., testing alignment of the quadrupole and octopole, of the lack of angular correlation at large angles) are mostly or solely sensitive to modes at $\ell<5$. We checked the impact of the chance correlations on a number of these test statistics in a similar way to Figure~\ref{fig:chisquare_hist}. These results are presented in  \cite{herold/etal:2025}. In general we find that the difference between the pure CMB and cleaned CMB histograms is small enough that the chance correlation impact can be neglected. The pure and cleaned CMB simulations are available for download so that the effect can be similarly quantified for any other tests as well \citep{Nofi:zenodo}.

\subsection{New $\ell<30$ power spectrum result}

The power spectrum and uncertainties of this paper are shown in Figure \ref{fig:cross_spectra_mean_errors}. 
The inner error bar is the quadruture sum of the foreground cleaning uncertainties shown in Figure~\ref{fig:cross_spectra} and the bias uncertainty associated with the blue points in Figure~\ref{fig:chance_corr_bias}. The outer error bar is the cosmic variance uncertainty that contains 68\% of the posterior inverse-gamma distribution (the probability of the underlying $C_{\ell}$ conditioned on the effectively full-sky measured power) at each 
multipole. Cosmic variance dominates at these low multipoles. 

The familiar low quadrupole and $\ell\sim 22$ dip are clearly seen, as previously reported by both WMAP and Planck.
Relative to the best-fit Planck 2018 \lcdm\ power spectrum (driven by high-$\ell$ measurements), we find from this new work that the quadrupole value is 
2.2$\sigma$ low,
and that the $\ell<30$ angular power spectrum presented here is lower by a factor of 0.92 determined using

\begin{equation}
\label{amplitude}
   \mathbf{ f_{A} = 
   \displaystyle \sum_{\ell=2}^{29}\left[{(2\ell +1) \frac{C_{\ell}^{data}}{C_{\ell}^{\Lambda CDM}}}\right]\Bigg /\displaystyle \sum_{\ell=2}^{29} (2\ell +1)}.
\end{equation}
based on the likelihood described in the next section.
Of our \lcdm\ simulations generated using the Planck best-fit model, only 3.5\%  had lower factors.

We compare the low-$\ell$ TT spectrum shown at the bottom of Figure~\ref{fig:cross_spectra} to those released by WMAP{\footnote{wmap\_tt\_spectrum\_9yr\_v5.txt}}
and Planck{\footnote{COM\_PowerSpect\_CMB-TT-full\_R3.01.txt}}.  Since these previous results were obtained using larger sky masks,
it is interesting to see what similarities and differences exist between these and our nearly full-sky determination.

The basis for the 9-yr WMAP $\ell \leq 32$ TT spectrum is a CMB map estimated from the de-biased Internal Linear Combination (ILC, \citealt{bennett/etal:2003c, bennett/etal:2013}) map degraded to $N_{\rm{side}}=32$.  The power spectrum is evaluated using the  KQ85y9 sky mask at the same resolution, leaving 77.3\% of the sky unmasked \citep{bennett/etal:2013}.

The low-$\ell$ ($\ell < 30$) TT power spectrum for both the PR3 and PR4 Planck releases is set to that derived using the
2018 {\texttt{Commander}} component separation performed using data from all nine Planck frequency bands \citep{tristram/etal:2024}. The CMB power spectrum is estimated using 86\% of the sky \citep{planck/05:2018, planck/06:2018}.

The published WMAP and Planck low-$\ell$ spectra are shown in Figure~\ref{fig:cross_spectra_overplot} as green and orange points respectively. The WMAP and Planck teams both used Gibbs sampling to estimate the posterior probability distribution of the $C_{\ell}$ given the observed maps and sky masks, under the assumptions of statistical isotropy and homogeneity, marginalizing over foreground and other nuisance parameters \citep{hinshaw/etal:2013, planck/05:2018}.

Diagonal uncertainties ($\pm1\sigma$) provided in the WMAP and Planck released files are also shown.  
However, the uncertainties for each mission are evaluated differently and also reflect different masking fractions.  
For WMAP, approximate error bars were provided for plotting purposes. These are symmetric and are dominated by a cosmic variance computed 
assuming $C_\ell$ from the WMAP best-fit \lcdm\ model.
The Planck uncertainties are asymmetric and include cosmic variance and uncertainty in the foreground model removal
as estimated from the \texttt{Commander} Gibbs samples.

The power spectrum from this work is represented in Figure~\ref{fig:cross_spectra_overplot} by the black points. There is generally good agreement between all three power spectra, despite the differences in sky fraction
and dramatically different analysis methods.  The close agreement between the points inspires confidence in
the robustness of the $\ell < 30$ power spectrum; we give further details in Section~\ref{subseclikelihood}.

\subsection{Likelihood}
\label{subseclikelihood}

We provide a power-spectrum based likelihood code for use in {\texttt{Cobaya}}, which may be substituted for the
existing $\ell<30$ WMAP or Planck low-$\ell$ likelihoods, and used in conjunction with Planck or other higher multipole CMB likelihoods.

We denote the measured power spectrum as $\hat{C}_{\ell}$ and the underlying theoretical power spectrum as $C_{\ell}$. For full-sky statistically isotropic and Gaussian fluctuations, the quantity $(2\ell+1)\hat{C}_{\ell}/C_{\ell}$ is distributed as a $\chi^2$ random variable with $2\ell+1$ degrees of freedom, with likelihood \citep[e.g.,][]{hamimeche/lewis:2008}
\be
P(\hat{C}_{\ell}|C_{\ell})\propto \frac{\hat{C}_{\ell}^{(2\ell-1)/2}}{C_{\ell}^{(2\ell+1)/2}}\exp\left(-\frac{(2\ell+1)\hat{C}_{\ell}}{2C_{\ell}}\right).
\ee

In addition to cosmic variance, our measured power spectra have some residual foreground uncertainty (estimated from the scatter in the cross-spectra shown in Fig.~\ref{fig:cross_spectra}). We modified the likelihood to include this extra error contribution, although it ultimately has a negligible impact on parameter constraints. We approximate the foreground uncertainty as an extra random Gaussian scatter on each measured $\hat{C}_{\ell}$. In reality these uncertainties will not be exactly Gaussian, however their effect at $\ell<30$ is so small that the exact treatment is unimportant.

Denoting this contribution to the measured power as $\delta_{\ell}$, the likelihood for the total measured power is then
\small
\be
 P(\hat{C}_{\ell}|C_{\ell}, \delta_{\ell})\propto \frac{(\hat{C}_{\ell}-\delta_{\ell})^{(2\ell-1)/2}}{C_{\ell}^{(2\ell+1)/2}}   \exp\left(-\frac{(2\ell+1)(\hat{C}_{\ell}-\delta_{\ell})}{2C_{\ell}}\right),
\ee
\normalsize
and we assume a Gaussian prior on $\delta_{\ell}$ with zero mean and standard deviation $\sigma_{\ell}$ given by the empirical standard deviation between the cross-spectra plotted in the bottom panel of Figure~\ref{fig:cross_spectra}.

Assuming that we are fitting for cosmological model parameters $\theta$, so that $C_{\ell}=C_{\ell}(\theta)$, the joint posterior for $\theta$ and $\delta_{\ell}$ using all measured multipoles can be written
\be
\begin{split}
P(\theta,\{\delta_{\ell}\}|\{\hat{C}_{\ell}\})\propto& P(\theta)\prod_{\ell=2}^{\ell_{\rm max}}\frac{(\hat{C}_{\ell}-\delta_{\ell})^{(2\ell-1)/2}}{C_{\ell}^{(2\ell+1)/2}}\\&\exp\left(-\frac{(2\ell+1)(\hat{C}_{\ell}-\delta_{\ell})}{2C_{\ell}(\theta)}-\frac{\delta_{\ell}^2}{2\sigma_{\ell}^2}\right),
\end{split}
\ee
where $P(\theta)$ is the prior on the parameters.

We found that when $\sigma_{\ell}/C_{\ell}$ is small (order 0.1 or less, as in this work), the posterior samples for $\delta_{\ell}$ closely follow the prior. Rather than sampling the $\delta_{\ell}$ as additional nuisance parameters in the parameter fitting, we therefore draw some large set of $\delta_{\ell}$ values from the prior once and estimate the marginalized posterior for $\theta$ at each chain step as
\be
\begin{split}
P(\theta|\{\hat{C_{\ell}}\})&=\int d^n\delta_{\ell}\,P(\theta,\{\delta_{\ell}\}|\{\hat{C}_{\ell}\})\\
&=\int d^n\delta_{\ell}\,P(\theta|\{\hat{C}_{\ell}\},\{\delta_{\ell}\})\,P(\{\delta_{\ell}\}|\hat{C}_{\ell}\})\\
&\approx\int d^n\delta_{\ell}\,P(\theta|\{\hat{C}_{\ell}\},\{\delta_{\ell}\})\,P(\{\delta_{\ell}\})\\
&=\left\langle P(\theta|\{\hat{C}_{\ell}\},\{\delta_{\ell}\})\right\rangle_{\delta_{\ell}\sim P(\delta_{\ell})}\\
&\approx\frac{1}{N_{\rm draw}}\sum_{i=1}^{N_{\rm draw}}P(\theta|\{\hat{C}_{\ell}\},\{\delta_{\ell,i}\}),
\end{split}
\ee
where $\delta_{\ell,i}$ denotes the values drawn from the prior and $N_{\rm draw}$ is the number of draws. To avoid unnecessary additional numerical noise we do not generate the $\delta_{\ell}$ values at random but use an even sampling of the Gaussian cumulative density function (CDF).

As an example of the results from this new likelihood, Table~\ref{table:As} shows constraints on the scalar amplitude, $A_s$, from WMAP and Planck data in the \lcdm\ model. We compare results from the full data sets (including temperature and polarization spectra, and the lensing reconstruction for Planck), with those from $\ell<30$ alone. For fits to the $\ell<30$ TT spectra alone we fix the other five \lcdm\ parameters to their best-fit values based on the Planck 2018 analysis. We follow the Planck team in sampling with a uniform prior on the quantity $\ln(10^{10}A_s)$, although the posterior distributions for both $\ln A_s$ and $A_s$ are well approximated as Gaussian.

The $A_s$ constraint from our new likelihood is around 1\% lower than that from the Planck 2018 $\ell<30$ TT spectrum, corresponding to 24\% of the statistical precision of our constraint. If we regard the Planck low-$\ell$ likelihood as containing a subset of the information contained in our analysis, which seems a reasonable assumption given the larger sky mask, the shift in the mean $A_s$ value has a statistical significance of $0.6\sigma$ based on the ``difference-of-covariance'' test used in the Planck papers \citep[e.g.,][]{planck/05:2018,gratton/challinor:2020}. The same test comparing our amplitude constraint with that from the WMAP-9 low-$\ell$ likelihood\footnote{A version of the WMAP likelihood that runs in \texttt{Cobaya} can be found at \url{https://github.com/HTJense/pyWMAP}.} yields a significance of $1.2\sigma$ given the change in constraining power. While some individual multipoles do visibly differ between the three spectra (Fig.~\ref{fig:cross_spectra_overplot}), the overall amplitude constraints  from our analysis are consistent with the previous results within statistical expectations, with no indication of significant unaccounted-for foreground residuals or other systematic biases.

We verified that the small difference in precision in $A_s$ between this work and Planck 2018 in Table~\ref{table:As} is consistent with the difference in sky fraction. From CMB-only simulations, we find a relative precision of $A_s$ of 4.7\% for the 1\% mask and $5.1-5.4$\% for a $10-20$\% mask. The Planck 2018 low-multipole mask is not publicly available but covers 14\% of the sky.

We also investigated the effect of using our $\ell<30$ TT likelihood in place of the Planck 2018 $\ell<30$ likelihood in joint fits with other Planck data (low-multipole polarization, high-multipole TT/TE/EE\footnote{Here, TE denotes the temperature-E-mode cross-power spectrum, and EE deonotes the E-mode polarization power spectrum.}, and lensing). In this case most of the weight comes from the higher multipoles and the parameter results are not meaningfully shifted or tightened. For example, in a \lcdm\ fit we find $H_0=67.35\pm0.54$~km~s$^{-1}$~Mpc$^{-1}$ with our likelihood, compared to $H_0=67.36\pm0.54$~km~s$^{-1}$~Mpc$^{-1}$ when using the Planck low-$\ell$ likelihood.

\begin{table}
  \caption{Posterior mean and 68\% intervals for scalar amplitude, $A_s$, from CMB measurements in the \lcdm\ model; other parameters are held fixed in $\ell<30$ TT fits}
   \label{table:As}
  \centering
  \begin{tabular}{llcc}
  \hline
  Analysis&Multipoles&$\ln(10^{10}A_s)$&$10^9A_s$\\
  \hline
  \hline
WMAP 9-year&Full&$3.091\pm0.031$&$2.203\pm0.068$\\
&$\ell<30$ TT&$2.990\pm0.051$&$1.990\pm0.102$\\
\hline
Planck 2018&Full&$3.044\pm0.014$&$2.100\pm0.030$\\
&$\ell<30$ TT&$2.949\pm0.051$&$1.911\pm0.098$\\
\hline
This work&$\ell<30$ TT&$2.961\pm0.047$&$1.933\pm0.091$
  \end{tabular}
\end{table}

\section{Conclusions}
\label{sec:conclusions}

\cite{nofi/etal:2025} recently provided foreground-cleaned CMB maps of 99\% of the sky at 1\degree resolution at 70, 94, 100, and 143 GHz, seen in Figure  \ref{fig:masked_cln_maps}. In this paper we used those maps to study the $\ell<30$ angular power spectrum.  We summarize our findings as follows:

(1) We took the 6 cross-spectra between pairs of those maps to further minimize foreground contamination and systematic errors.  WMAP and Planck instrument noise levels are generally minor at these scales. The resulting angular power spectra are shown in Figure \ref{fig:cross_spectra}. The different frequency cross-spectra are in excellent agreement. 

(2) Using simulations, we compute a potential bias correction due to chance correlations between the CMB and foreground templates. However, we find that if we apply a mean bias correction to the
cleaned spectra, it does not lead to better recovery.  The bias is greatest for $\ell \lesssim 5$, but in practice
the difference in the $\chi^2$ distributions per $\ell$-mode between simulated pure and cleaned CMB maps is negligible.
Therefore, we conclude a bias correction for chance correlations is not needed, as seen in Figure~\ref{fig:chisquare_hist}.

(3) The derived angular power spectrum generally reproduces previous results (see Figure \ref{fig:cross_spectra_overplot}), but uses a completely different methodology with only 1\% of pixels masked ($f_{sky}=0.99$), essentially eliminating mode-mixing. The WMAP 9yr and Planck PR3 analyses used 77\% and 86\% of the sky respectively. 

(4) Notable features continue to appear: (a) low quadrupole power compared to the best-fit Planck 2018 \lcdm\ spectrum by $2.2\sigma$, (b) a dip in the range $20\leq \ell \leq 27$, and (c) an overall $\ell<30$ power level low of the $\Lambda$CDM prediction derived from higher multipole moments.  
\emph{We conclude that there is essentially no chance that these features arise from foregrounds, systematic errors, masking, or mode-mixing.} 

(5) We find the $\ell<30$ angular power spectrum presented here prefers an overall amplitude rescaling factor of 0.92 relative to the full Planck 2018 \lcdm\ best-fit model. Of our \lcdm\ simulations, 3.5\% had lower power factors.

(6) We provide a new low-$\ell$ TT cosmological likelihood that can be used in the Cobaya MCMC package as an alternative for low-$\ell$ TT likelihoods released by WMAP and Planck teams. We illustrate its use in subsection \ref{subseclikelihood}.

(7) While some individual multipoles in our power spectrum visibly differ from the published WMAP and Planck results, the constraints on the overall $\ell<30$ amplitude are consistent within statistical expectations given the slightly sharper precision of our new likelihood (due to the larger sky fraction), with no indication of the presence of significant foreground contamination, etc. Accounting for the change in statistical precision, our amplitude result is consistent with WMAP at $1.2\sigma$ and Planck at $0.6\sigma$.

The pure and cleaned CMB simulations, the low-multipole TT cosmological likelihood, and the code to reproduce the results in this work are available in a Zenodo repository \citep{Nofi:zenodo}.

\vspace*{0.25in}

This research was supported by NASA grants 80NSSC23K0475, 80NSSC24K0625, and 80NSSC25K7518.
We acknowledge the use of the Legacy Archive for Microwave Background Data Analysis (LAMBDA), part of the High Energy Astrophysics Science Archive Center (HEASARC). HEASARC/LAMBDA is a service of the Astrophysics Science Division at the NASA Goddard Space Flight Center.
We also acknowledge the use of the Planck Legacy Archive. Planck is an ESA science mission with instruments and contributions
directly funded by ESA Member States, NASA, and Canada.

\software{numpy \citep{harris/etal:2020}, scipy \citep{virtanen/etal:2020}, matplotlib \citep{hunter:2007}, astropy \citep{astropy2022}, HEALPix \citep{gorski/etal:2005},
HEALPy \citep{healpy:2019}, Cobaya \citep{cobaya:2021}}

\bibliographystyle{aasjournal}
\bibliography{main,wmap,software,janet_planck}

\begin{thebibliography}{}
\expandafter\ifx\csname natexlab\endcsname\relax\def\natexlab#1{#1}\fi
\providecommand{\url}[1]{\href{#1}{#1}}
\providecommand{\dodoi}[1]{doi:~\href{http://doi.org/#1}{\nolinkurl{#1}}}
\providecommand{\doeprint}[1]{\href{http://ascl.net/#1}{\nolinkurl{http://ascl.net/#1}}}
\providecommand{\doarXiv}[1]{\href{https://arxiv.org/abs/#1}{\nolinkurl{https://arxiv.org/abs/#1}}}

\bibitem[{{Astropy Collaboration} {et~al.}(2022){Astropy Collaboration}, {Price-Whelan}, {Lim}, {Earl}, {Starkman}, {Bradley}, {Shupe}, {Patil}, {Corrales}, {Brasseur}, {N{\"o}the}, {Donath}, {Tollerud}, {Morris}, {Ginsburg}, {Vaher}, {Weaver}, {Tocknell}, {Jamieson}, {van Kerkwijk}, {Robitaille}, {Merry}, {Bachetti}, {G{\"u}nther}, {Aldcroft}, {Alvarado-Montes}, {Archibald}, {B{\'o}di}, {Bapat}, {Barentsen}, {Baz{\'a}n}, {Biswas}, {Boquien}, {Burke}, {Cara}, {Cara}, {Conroy}, {Conseil}, {Craig}, {Cross}, {Cruz}, {D'Eugenio}, {Dencheva}, {Devillepoix}, {Dietrich}, {Eigenbrot}, {Erben}, {Ferreira}, {Foreman-Mackey}, {Fox}, {Freij}, {Garg}, {Geda}, {Glattly}, {Gondhalekar}, {Gordon}, {Grant}, {Greenfield}, {Groener}, {Guest}, {Gurovich}, {Handberg}, {Hart}, {Hatfield-Dodds}, {Homeier}, {Hosseinzadeh}, {Jenness}, {Jones}, {Joseph}, {Kalmbach}, {Karamehmetoglu}, {Ka{\l}uszy{\'n}ski}, {Kelley}, {Kern}, {Kerzendorf}, {Koch}, {Kulumani}, {Lee}, {Ly}, {Ma}, {MacBride}, {Maljaars}, {Muna}, {Murphy}, {Norman},
  {O'Steen}, {Oman}, {Pacifici}, {Pascual}, {Pascual-Granado}, {Patil}, {Perren}, {Pickering}, {Rastogi}, {Roulston}, {Ryan}, {Rykoff}, {Sabater}, {Sakurikar}, {Salgado}, {Sanghi}, {Saunders}, {Savchenko}, {Schwardt}, {Seifert-Eckert}, {Shih}, {Jain}, {Shukla}, {Sick}, {Simpson}, {Singanamalla}, {Singer}, {Singhal}, {Sinha}, {Sip{\H{o}}cz}, {Spitler}, {Stansby}, {Streicher}, {{\v{S}}umak}, {Swinbank}, {Taranu}, {Tewary}, {Tremblay}, {de Val-Borro}, {Van Kooten}, {Vasovi{\'c}}, {Verma}, {de Miranda Cardoso}, {Williams}, {Wilson}, {Winkel}, {Wood-Vasey}, {Xue}, {Yoachim}, {Zhang}, {Zonca}, \& {Astropy Project Contributors}}]{astropy2022}
{Astropy Collaboration}, {Price-Whelan}, A.~M., {Lim}, P.~L., {et~al.} 2022, \apj, 935, 167, \dodoi{10.3847/1538-4357/ac7c74}

\bibitem[{{Bennett} {et~al.}(1996){Bennett}, {Banday}, {G{\'o}rski}, {Hinshaw}, {Jackson}, {Keegstra}, {Kogut}, {Smoot}, {Wilkinson}, \& {Wright}}]{bennett/etal:1996}
{Bennett}, C.~L., {Banday}, A.~J., {G{\'o}rski}, K.~M., {et~al.} 1996, \apjl, 464, L1

\bibitem[{{Bennett} {et~al.}(2003){Bennett}, {Hill}, {Hinshaw}, {Nolta}, {Odegard}, {Page}, {Spergel}, {Weiland}, {Wright}, {Halpern}, {Jarosik}, {Kogut}, {Limon}, {Meyer}, {Tucker}, \& {Wollack}}]{bennett/etal:2003c}
{Bennett}, C.~L., {Hill}, R.~S., {Hinshaw}, G., {et~al.} 2003, \apjs, 148, 97

\bibitem[{{Bennett} {et~al.}(2011){Bennett}, {Hill}, {Hinshaw}, {Larson}, {Smith}, {Dunkley}, {Gold}, {Halpern}, {Jarosik}, {Kogut}, {Komatsu}, {Limon}, {Meyer}, {Nolta}, {Odegard}, {Page}, {Spergel}, {Tucker}, {Weiland}, {Wollack}, \& {Wright}}]{bennett/etal:2011}
---. 2011, \apjs, 192, 17, \dodoi{10.1088/0067-0049/192/2/17}

\bibitem[{{Bennett} {et~al.}(2013){Bennett}, {Larson}, {Weiland}, {Jarosik}, {Hinshaw}, {Odegard}, {Smith}, {Hill}, {Gold}, {Halpern}, {Komatsu}, {Nolta}, {Page}, {Spergel}, {Wollack}, {Dunkley}, {Kogut}, {Limon}, {Meyer}, {Tucker}, \& {Wright}}]{bennett/etal:2013}
{Bennett}, C.~L., {Larson}, D., {Weiland}, J.~L., {et~al.} 2013, \apjs, 208, 20, \dodoi{10.1088/0067-0049/208/2/20}

\bibitem[{{Delouis} {et~al.}(2019){Delouis}, {Pagano}, {Mottet}, {Puget}, \& {Vibert}}]{delouis/etal:2019}
{Delouis}, J.~M., {Pagano}, L., {Mottet}, S., {Puget}, J.~L., \& {Vibert}, L. 2019, \aap, 629, A38, \dodoi{10.1051/0004-6361/201834882}

\bibitem[{Gorski {et~al.}(2005)Gorski, Hivon, Banday, Wandelt, Hansen, Reinecke, \& Bartlemann}]{gorski/etal:2005}
Gorski, K.~M., Hivon, E., Banday, A.~J., {et~al.} 2005, \apj, 622, 759

\bibitem[{{Gratton} \& {Challinor}(2020)}]{gratton/challinor:2020}
{Gratton}, S., \& {Challinor}, A. 2020, \mnras, 499, 3410, \dodoi{10.1093/mnras/staa2996}

\bibitem[{{Hamimeche} \& {Lewis}(2008)}]{hamimeche/lewis:2008}
{Hamimeche}, S., \& {Lewis}, A. 2008, \prd, 77, 103013, \dodoi{10.1103/PhysRevD.77.103013}

\bibitem[{Harris {et~al.}(2020)Harris, Millman, van~der Walt, Gommers, Virtanen, Cournapeau, Wieser, Taylor, Berg, Smith, Kern, Picus, Hoyer, van Kerkwijk, Brett, Haldane, del R{\'{i}}o, Wiebe, Peterson, G{\'{e}}rard-Marchant, Sheppard, Reddy, Weckesser, Abbasi, Gohlke, \& Oliphant}]{harris/etal:2020}
Harris, C.~R., Millman, K.~J., van~der Walt, S.~J., {et~al.} 2020, Nature, 585, 357, \dodoi{10.1038/s41586-020-2649-2}

\bibitem[{{Herold} {et~al.}(2025){Herold}, {Addison}, {Bennett}, {Nofi}, \& {Weiland}}]{herold/etal:2025}
{Herold}, L., {Addison}, G.~E., {Bennett}, C.~L., {Nofi}, H.~C., \& {Weiland}, J.~L. 2025, {in prep}

\bibitem[{{Hinshaw} {et~al.}(1996){Hinshaw}, {Banday}, {Bennett}, {G{\'o}rski}, {Kogut}, {Smoot}, \& {Wright}}]{hinshaw/etal:1996a}
{Hinshaw}, G., {Banday}, A.~J., {Bennett}, C.~L., {et~al.} 1996, \apjl, 464, L17

\bibitem[{{Hinshaw} {et~al.}(2013){Hinshaw}, {Larson}, {Komatsu}, {Spergel}, {Bennett}, {Dunkley}, {Nolta}, {Halpern}, {Hill}, {Odegard}, {Page}, {Smith}, {Weiland}, {Gold}, {Jarosik}, {Kogut}, {Limon}, {Meyer}, {Tucker}, {Wollack}, \& {Wright}}]{hinshaw/etal:2013}
{Hinshaw}, G., {Larson}, D., {Komatsu}, E., {et~al.} 2013, \apjs, 208, 19, \dodoi{10.1088/0067-0049/208/2/19}

\bibitem[{{Hivon} {et~al.}(2002){Hivon}, {G{\' o}rski}, {Netterfield}, {Crill}, {Prunet}, \& {Hansen}}]{hivon/etal:2002}
{Hivon}, E., {G{\' o}rski}, K.~M., {Netterfield}, C.~B., {et~al.} 2002, \apj, 567, 2

\bibitem[{Hunter(2007)}]{hunter:2007}
Hunter, J.~D. 2007, Computing in Science \& Engineering, 9, 90, \dodoi{10.1109/MCSE.2007.55}

\bibitem[{{Nofi} {et~al.}(2025){Nofi}, {Addison}, {Bennett}, {Herold}, \& {Weiland}}]{nofi/etal:2025}
{Nofi}, H.~C., {Addison}, G.~E., {Bennett}, C.~L., {Herold}, L., \& {Weiland}, J.~L. 2025, {in prep}

\bibitem[{{Planck Collaboration I}(2020)}]{planck/01:2018}
{Planck Collaboration I}. 2020, \aap, 641, A1, \dodoi{10.1051/0004-6361/201833880}

\bibitem[{{Planck Collaboration Int. LVII}(2020)}]{npipe:2020}
{Planck Collaboration Int. LVII}. 2020, \aap, 643, A42, \dodoi{10.1051/0004-6361/202038073}

\bibitem[{{Planck Collaboration V}(2020)}]{planck/05:2018}
{Planck Collaboration V}. 2020, \aap, 641, A5, \dodoi{10.1051/0004-6361/201936386}

\bibitem[{{Planck Collaboration VI}(2020)}]{planck/06:2018}
{Planck Collaboration VI}. 2020, \aap, 641, A6, \dodoi{10.1051/0004-6361/201833910}

\bibitem[{{Planck Collaboration VII}(2020)}]{planck/07:2018}
{Planck Collaboration VII}. 2020, \aap, 641, A7, \dodoi{10.1051/0004-6361/201935201}

\bibitem[{{\sorthelp{Planck Collaboration 2018D}}{Planck Collaboration IV}(2020)}]{planck/04:2018}
{\sorthelp{Planck Collaboration 2018D}}{Planck Collaboration IV}. 2020, \aap, 641, A4, \dodoi{10.1051/0004-6361/201833881}

\bibitem[{{Schwarz} {et~al.}(2016){Schwarz}, {Copi}, {Huterer}, \& {Starkman}}]{schwarz/etal:2016}
{Schwarz}, D.~J., {Copi}, C.~J., {Huterer}, D., \& {Starkman}, G.~D. 2016, Classical and Quantum Gravity, 33, 184001, \dodoi{10.1088/0264-9381/33/18/184001}

\bibitem[{{Tegmark}(1997)}]{tegmark:1997}
{Tegmark}, M. 1997, \prd, 55, 5895

\bibitem[{{Torrado} \& {Lewis}(2021)}]{cobaya:2021}
{Torrado}, J., \& {Lewis}, A. 2021, \jcap, 2021, 057, \dodoi{10.1088/1475-7516/2021/05/057}

\bibitem[{{Tristram} {et~al.}(2024){Tristram}, {Banday}, {Douspis}, {Garrido}, {G{\'o}rski}, {Henrot-Versill{\'e}}, {Hergt}, {Ili{\'c}}, {Keskitalo}, {Lagache}, {Lawrence}, {Partridge}, \& {Scott}}]{tristram/etal:2024}
{Tristram}, M., {Banday}, A.~J., {Douspis}, M., {et~al.} 2024, \aap, 682, A37, \dodoi{10.1051/0004-6361/202348015}

\bibitem[{Virtanen {et~al.}(2020)Virtanen, Gommers, Oliphant, Haberland, Reddy, Cournapeau, Burovski, Peterson, Weckesser, Bright, {van der Walt}, Brett, Wilson, Millman, Mayorov, Nelson, Jones, Kern, Larson, Carey, Polat, Feng, Moore, {VanderPlas}, Laxalde, Perktold, Cimrman, Henriksen, Quintero, Harris, Archibald, Ribeiro, Pedregosa, {van Mulbregt}, \& {SciPy 1.0 Contributors}}]{virtanen/etal:2020}
Virtanen, P., Gommers, R., Oliphant, T.~E., {et~al.} 2020, Nature Methods, 17, 261, \dodoi{10.1038/s41592-019-0686-2}

\bibitem[{Zonca {et~al.}(2019)Zonca, Singer, Lenz, Reinecke, Rosset, Hivon, \& Gorski}]{healpy:2019}
Zonca, A., Singer, L., Lenz, D., {et~al.} 2019, Journal of Open Source Software, 4, 1298, \dodoi{10.21105/joss.01298}

\end{thebibliography}

\end{document}